\documentclass[reprint,amsmath,amssymb,aps,floatfix,]{revtex4-2}

\usepackage{graphicx}
\usepackage{dcolumn}
\usepackage{bm}
\usepackage{subcaption}
\usepackage{ragged2e}
\usepackage{url}
\usepackage{hyperref}
\usepackage[dvipsnames]{xcolor}

\begin{document}
\title{\textbf{Time Series Extrinsic Regression of Ion Cyclotron Emission Spectra Trained on Particle-In-Cell Simulations}
}%

\author{Ethan Attwood}
 \email{Contact author: ethan.attwood@york.ac.uk}
\affiliation{%
 York Plasma Institute, University of York, York, UK}
\author{J. W. S. Cook}
 \email{james.cook@ukaea.uk}
\affiliation{
 UK Atomic Energy Authority, Culham Campus, Culham, UK
}%
\affiliation{Centre for Fusion, Space and Astrophysics, University of Warwick, Coventry, CV4 7AL, United Kingdom of Great Britain and Northern Ireland}
\author{Peter Hill}%
 \email{peter.hill@york.ac.uk}
\affiliation{%
 York Plasma Institute, University of York, York, UK}

\date{\today}

\begin{abstract}
Ion Cyclotron Emission (ICE) is a ubiquitous magnetised plasma phenomenon previously detected on virtually all large magnetic fusion devices and whose diagnostic potential for future power plants rests upon an accurate mapping of plasma parameters to spectra. This work demonstrates that the inverse problem is solved by training Time-Series Extrinsic Regression (TSER) models on synthetic ICE spectra from oblique propagation angle sweeps of nonlinear fully kinetic 1D3V particle-in-cell simulations of the magnetoacoustic cyclotron instability. Using datasets from a systematically constructed scan over reactor-relevant ranges of background magnetic field strength, density, and alpha-particle velocity pitch ($v_\parallel/|v|$) and concentration, we show that these bulk and fast ion parameters may be recovered from a JET ICE spectrum via TSER models with near real-time capability.
\end{abstract}

\maketitle
\newpage

Magnetically-confined fusion devices require a network of diagnostics to determine plasma properties critical to safe and effective operation. Next-generation devices increasingly demand diagnostics that maximize information efficiency and wall space for tritium breeding through the surrounding lithium blanket. For example, nearly 20\% of the outboard vessel area of ITER is used for diagnostics, damaging the tritium-breeding ratio, which is critical for the transition from research-scale to grid-scale fusion devices~\cite{todd2014diagnostic}. Diagnostic solutions demonstrating a high technical readiness and providing data on a wide variety of plasma characteristics are more important than ever.

Ion Cyclotron Emission (ICE) is a widespread plasma phenomenon, occurring in virtually all large hot magnetized plasmas \cite{cottrell1988superthermal, cottrell1993ion, fulop1998ion, sato2010observation, dendy2015ion, mcclements2015fast, gorelenkov2016ion, mcclements2018observations, ochoukov2018observations, askinazi2018ion, thome2019central, liu2020ion} and experimentally observed as suprathermal emission with spectral peaks at multiple harmonics of the ion cyclotron frequency. Typically, ICE requires a minority fast ion population in cyclotron resonance with a fast Alfv\'en wave propagating approximately perpendicular to the background magnetic field, which gives rise to the magnetoacoustic cyclotron instability (MCI), the leading mechanism underpinning ICE~\cite{dendy1992possible, dendy1995ion, fulop1997origin, carbajal2014linear, gorelenkov2016ion, ochoukov2019interpretation, chapman2020origin, slade2025effects}. An abundant source of fast ions in a tokamak is the fusion reaction itself, thereby correlating ICE observations with fusion reaction rate as well as with the magnetic field strength and the spatial and velocity distribution of energetic ion populations via the finer details of the kinetic instability~\cite{cottrell1993ion, mcclements1996interpretation}. Ion cyclotron resonance heated\cite{mcclements2018observations} and neutral beam injection ions\cite{reman2019interpreting} are also known to excite ICE.

The Particle-In-Cell (PIC) method~\cite{hockney1988computer, birdsall1991plasma,Arber:2015hc} is a common technique in computational plasma physics for self-consistently solving Maxwell's equations on a grid simultaneously with currents and charge densities provided by plasma macroparticles subject to the Lorentz force. PIC is ideal for computationally efficient models of nonlinear kinetic phenomena evolving at very fast timescales such as ion gyroperiods, and has been widely applied to ICE investigations~\cite{cook2013particle, carbajal2014linear, gorelenkov2016ion, reman2019interpreting, chapman2020comparing, dendy2023mechanism}. Resolution constraints placed on the number and size of cells, particle numbers and simulation durations make large parameter scans with PIC simulations computationally expensive albeit tractable. Reasonable simplifications can be made such as lowering dimensionality, which we apply in this work by rendering our problem in so-called 1D3V: one spatial dimension of variation is allowed while vector fields, including velocities, retain three components. In this letter, we exploit a fully kinetic approach for all species to capture electron Landau damping with non-zero parallel wavenumber behaviour and ion gyro-resonances crucial to accurately modelling the MCI~\cite{cook2022doublet}. In order to include some two dimensional physics, we run four 1D3V simulations for each set of parameters: one perpendicular to the magnetic field, $90^\circ$, and three at oblique angles, $92^\circ$, $94^\circ$ and $96^\circ$. This retains one dimensional nonlinear physics and introduces linear two dimensional physics at a substantially lower computational cost than 2D3V simulations. Experimental ICE measurements will tend to reflect the nonlinear regime of the MCI due to its dominance past an initial single-digit number of ion gyroperiods~\cite{carbajal2014linear} as well as the complexities of the machine geometry, wave propagation, and sink and source terms. Fortunately, analytic approaches and periodic 1D3V PIC code simulations that mimic the infinite homogeneous plasma limit have been able to explain many aspects of ICE, including how the spectral features that arise during the linear phase of the instability leave their imprint on the nonlinear spectrum~\cite{cook2013particle, carbajal2014linear, reman2019interpreting, chapman2020comparing}. Nonlinearity, enabled by PIC, is required in order to account for the magnitude of energy transfer from fast ions and for wave-wave coupling of power from higher to lower harmonics\cite{chapman2018nonlinear} and hence is essential for comparing simulation results to experimental observations.

We performed parameter scans to form ICE-relevant datasets, on which machine learning models were trained to infer meaningful physical properties. Parameter sets can be designed specifically for individual magnetised fusion devices, a natural first step in the development of synthetic diagnostics. We chose JET-like parameters due to the availability of ICE observations from tritium experiments with a high drive of fusion alpha-particles which, as opposed to lower energy neutral injection ions, feature increased ICE intensity for this first of a kind study.

The dataset was exploited by regressing spectra against the input simulation parameters. Regression of such sequential data (known in machine learning literature as ``time-series'' data; whether the dynamics are temporal or based on some other sequential ordering is immaterial) against continuous external values is a young field, having only been formalised by Tan et al. in 2021~\cite{tan2021time}. ICE signals derived from simulations are power spectra calculated by Fourier transforming temporal field data, which we call synthetic spectra as we use them as analogues for experimentally observed spectra. We applied this approach to ICE where we trained models capable of determining the following four quantities from synthetic ICE spectra: the equilibrium magnetic field strength ($B_0$); electron density ($n_e$); concentration of fusion-born alpha-particles ($n_\alpha/n_e$); and alpha-particle velocity distribution pitch $\lambda ={u_{\parallel}}/\sqrt{(u_\parallel^2 + u_\perp^2)}$, defined as the ratio of parallel bulk speed to the initial speed, which determines their distribution function in Equation~\ref{eqn:df}.

In our simulations, three species were modelled: electrons and background fuel deuterons, both initially at 10kev, and a minority of energetic alpha particles approximated as a ``ring-beam'' delta function,
\begin{equation}
    f(v_\parallel, v_\perp) \propto \exp\left(-\frac{(v_\parallel - u_\parallel)^2}{{v_{th,\parallel}^2}} -\frac{(v_\perp - u_\perp)^2}{{v_{th,\perp}^2}}\right),
\label{eqn:df}
\end{equation}
\noindent which initialises particle velocities according to bulk speeds parallel and perpendicular to the magnetic field direction, $u_\parallel$ and $u_{\perp}$ respectively, and Maxwellian thermal speeds $v_{th,\parallel}$ and $v_{th,\perp}$. An homogeneous magnetic field was initialised predominantly in $z$, perpendicular to the spatial domain $x$, such that the MCI is most strongly excited~\cite{gorelenkov1995alfven}.

To generate the datasets, four quantities were sampled using a Latin hypercube strategy, the minimum and maximum values of which are shown in Table~\ref{tab:epoch_vars}. Electron density and alpha-particle fraction were sampled uniformly in log space. Static parameters used in all simulations are listed in Table~\ref{tab:static_params}. Initially 100 parameter sets were simulated; later 10 more were added using another Latin hypercube near to the known parameter values of JET pulse \#26148 ($B_0=2-3$T, $\lambda=0.3-0.7$, $n_e=(1-2)\times10^{19}\mathrm{m}^{-3}$, $n_\alpha/n_e=10^{-4}-10^{-3}$) as documented in Ref.~\cite{cottrell1993ion}. Reference machine values are shown in the rightmost column for $B_0$ and $n_e$ in Table~\ref{tab:epoch_vars}. As ICE originates from a variety of fast ion sources (NBI, ICRH, fusion reactions) in both the tokamak core\cite{ochoukov2018observations, ochoukov2019interpretation} and edge plasma regions~\cite{chapman2020comparing}, there are a number of valid operational combinations of parameters such as density and species temperature. This dataset trades specificity to one particular regime for a versatile training set representative of a broad sample of ICE origins.
\begin{table}[h!]
    \begin{ruledtabular}
        \begin{tabular}{l|llll|lll}
            \textrm{Parameter}&
            \textrm{Min}&
            \textrm{Max}&
            \textrm{Mean}&
            \textrm{SD}&
            \textrm{JET}&
            \textrm{ITER}&
            \textrm{STEP}\\
            \colrule
            $B_0$ [T] & 0.5 & 5.0 & $2.75$ & $1.16$ & 3.4 & 5.3& 3.2\\
            $\lambda\,(v_\parallel/|v|)$ & 0.01 & 0.99 & $0.500$ & $0.283$ &--&--&--\\
            $n_e$ [$\log_{10}(\mathrm{m}^{-3})$] & $19$ & $20$ & $19.5$ & $0.289$ &19.7&19.5&20.0\\
            $n_\alpha/n_e$ [$\log_{10}$] & $-4$ & $-2$ & $-3.00$ & $0.577$ &--&--&--\\
        \end{tabular}
    \end{ruledtabular}
    \caption{Simulation parameter scan variable ranges, means and standard deviations (SD). Electron density ($n_e$) and alpha-particle fraction ($n_\alpha/n_e$) were sampled uniformly in $\log_{10}$ space. The rightmost columns show typical values of $B$ (toroidal, on-axis) and $n_e$ in previous D-T experiments on JET~\cite{gormezano2008chapter}, or projected for ITER~\cite{Sips_2005} and STEP~\cite{Tholerus_2024}. Densities are averages across major radius and operational configurations. Minimum and maximum values are shown at the precision to which they were specified; mean and SD's were calculated post-sampling and shown to 3 significant figures.}
    \label{tab:epoch_vars}
\end{table}
\begin{table}[h!]
    \begin{ruledtabular}
        \begin{tabular}{l|l|l}
            \textrm{Symbol}&
            \textrm{Description}&
            \textrm{Value}\\
            \colrule
            $T_{D,e}$ & background temperature [eV] & 10000 \\
            $E_\alpha$ & $\alpha$ particle energy [eV] & 3500000 \\
            $p_\alpha$ & $\alpha$ particle momentum [kgms$^{-1}$] & $\sqrt{2m_\alpha q_e E_\alpha}$ \\
            $p_{\text{beam}}$ & $(u_\parallel m_\alpha)$ [kgms$^{-1}$] & $p_\alpha \lambda$ \\ 
            $p_{\text{ring}}$ & $(u_\perp m_\alpha)$ [kgms$^{-1}$] & $p_\alpha \sqrt{1 - \lambda^2}$  \\ 
            $p_{\text{spread}}$ & $(v_{th}m_{\alpha})$ [kgms$^{-1}$] & $0.06 p_{\text{ring}}$ \\
            $t_{\text{sim}}$ & simulation time [$\tau_{c,\alpha}$] & 16 \\
            $N_{\Omega}$ & Nyquist frequency [$\Omega_{c,\alpha}$] & 80 \\
            $n_{\text{samples}}$ & number of time samples & $2 t_{\text{sim}} N_\Omega$ \\
            $n_{\text{ppc}}$ & particles per species per cell & 250 \\
        \end{tabular}
    \end{ruledtabular}
    \caption{Static parameters used in all simulations. Times and frequencies were normalised to alpha-particle gyromotion ($\tau_{c,\alpha}$ and $\Omega_{c,\alpha}$ respectively).}
    \label{tab:static_params}
\end{table}

All simulations were run on ARCHER2~\cite{beckett_2024_14507040}, the UK Tier 1 HPC resource, with 250 macroparticles per species per cell for 16 deuteron gyroperiods, with a diagnostic output rate allowing resolution of data in frequency space up to 80 deuteron gyrofrequencies, which exceeds the frequency range of common experimental ICE spectra\cite{cottrell1993ion, reman2019interpreting}. Cell widths were set to the smallest of the electron thermal gyroradius or the Debye length for each set of parameters, indicated by previous studies~\cite{chapman2020origin, slade2024consequences} to benefit energy conservation. EPOCH default settings were used except for the enabling of $\delta f$, a technique used to suppress numerical noise~\cite{Arber:2015hc, lanti2020orb5}, and higher-order particle shape functions (a third-order b-spline)~\cite{Arber:2015hc}. A comparison of a subset of particle and field energy diagnostics with and without using $\delta f$ for the background deuterons and electrons showed no appreciable impact on the rate or magnitude of MCI evolution. A full-$f$ treatment was preserved for the fast alpha population. Running all simulations combined took approximately 1.55M CPU hours. Analysis of the outputs common to all simulations took approximately 600 CPU hours split across ARCHER2 and Viking, the University of York's compute cluster. This relatively modest level of computational resource is within reach of the majority of fusion research groups.

\begin{figure}[h!]
    \centering
    \includegraphics[width=\linewidth]{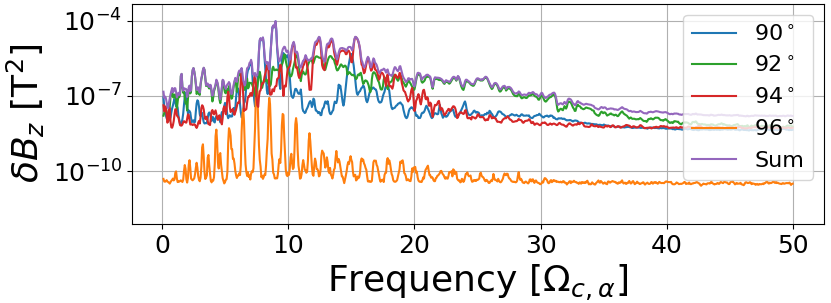}
    \caption{Frequency power spectra in $\delta B_z$ for four propagation angles of parameter set 75. Successive peaks are seen at integer harmonics of the deuteron/alpha cyclotron frequency, with peaks at $8-16\Omega_{c,\alpha}$ (frequency normalised to the alpha-particle cyclotron frequency), indicative of MCI activity. All four angles share a region of activity below $12\Omega_{c,\alpha}$. The perpendicular case, $90^\circ$ shows two distinct regions of excitation around $8\Omega_{c,\alpha}$ and $15\Omega_{c,\alpha}$ and the $96^\circ$ spectral amplitudes are 1000 times less powerful. The parameters are $B_0=2.0$T , $\lambda=0.49$, $n_e=1.4\times10^{19}$ and $n_\alpha=2.8\times10^{-3}n_e$.}
    \label{fig:freqSpec_75}
\end{figure}

The experimentally observed power spectra considered here and our synthetic spectra are a function of frequency only; all wavenumber and propagation information is lost. The electromagnetic field perturbations (parallel magnetic field $\delta B_z$, electrostatic component $\delta E_x$ and Alfv{\'e}nic electric component $\delta E_y$) are integrated over one-dimensional wavenumbers and summed across all four propagation angles of our 110 cases. A representative spectrum of the compressional magnetic field component $\delta B_z$, which provides the principle analogue for the MCI as measured by a magnetic field probe, is shown in Figure~\ref{fig:freqSpec_75} and illustrates that the dominant spectral features can arise from different propagation angles.
Trends in peak grouping, splitting, and amplitude with variation in propagation angle do not vary straightforwardly with the parameters. The fast magnetoacoustic dispersion surface on the parallel and perpendicular wavenumber plane is controlled largely by density and magnetic field strength. However, the complicated and sometimes disconnected regions of instability on this surface are determined further by the details of the fast ion distribution function and the damping arising from the temperatures of electrons and fuel ions: see Fig. 2 of Ref. \cite{cook2022doublet}. The mapping from spectra to these physics parameters includes details of these complex linear and nonlinear physical processes, which necessarily must be encapsulated by well-performing regression models. 

Time series analysis is the field of performing machine learning based on a sequence of values whose order and magnitude relative to the rest of the series is significant~\cite{tan2021time}. Time Series Extrinsic Regression (TSER) is a novel approach in plasma physics, having to our knowledge only once previously been applied to the reconstruction of missing electron temperature data~\cite{wang2025time}. We use the TSER~\cite{bagnall2024hands, guijo2024unsupervised, middlehurst2023extracting} capabilities of the Aeon time series analysis toolkit~\cite{JMLR:v25:23-1444} to apply these algorithms to the novel area of ICE parameter inference.

\begin{figure}[h!]
    \centering
    \includegraphics[width=\linewidth]{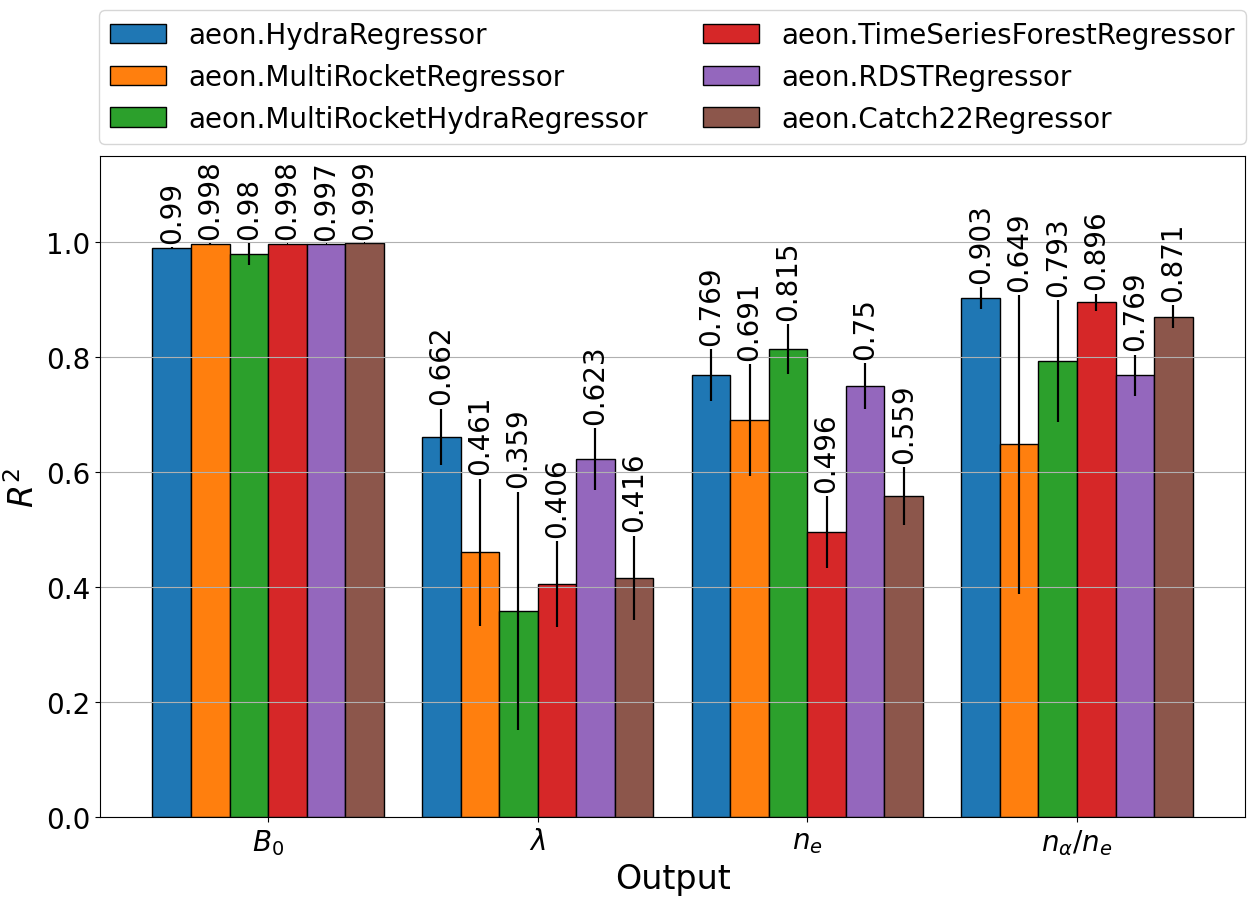}
    \caption{TSER results of a leave-one-out cross-validation of six algorithms against four simulation parameters. Highest accuracy algorithms in LOOCV and experimental comparison were selected based on the coefficient of determination (higher is better) across all predictions. Error bars show the mean standard error of prediction across all folds of the LOOCV.}
    \label{fig:tser_r2_bar}
\end{figure}

Each TSER model was trained on four input channels ($\log_{10}$ of the three EM spectra on a frequency range normalized by the per-case gyrofrequency; the fourth channel mapped gyrofrequencies to Hz) and regressed against a single output channel at a time (each of the plasma parameters in Table~\ref{tab:epoch_vars}). Cross-validation strategies are used in machine learning research to mitigate against over-fitting during model training. This involves partitioning the training data into folds and omitting a subset of the data, which is subsequently used to test the model. We use a leave-one-out cross-validation (LOOCV) strategy, where only one case is left out and reserved for testing, to maximise utility of our computationally expensive nonlinear simulations. Error metrics are evaluated over all LOOCV folds to generate holistic measures of model performance. Figure~\ref{fig:tser_r2_bar} shows the coefficient of determination ($R^2$) across predictions for every fold.

\begin{figure}[h!]
    \centering
    \includegraphics[width=\linewidth]{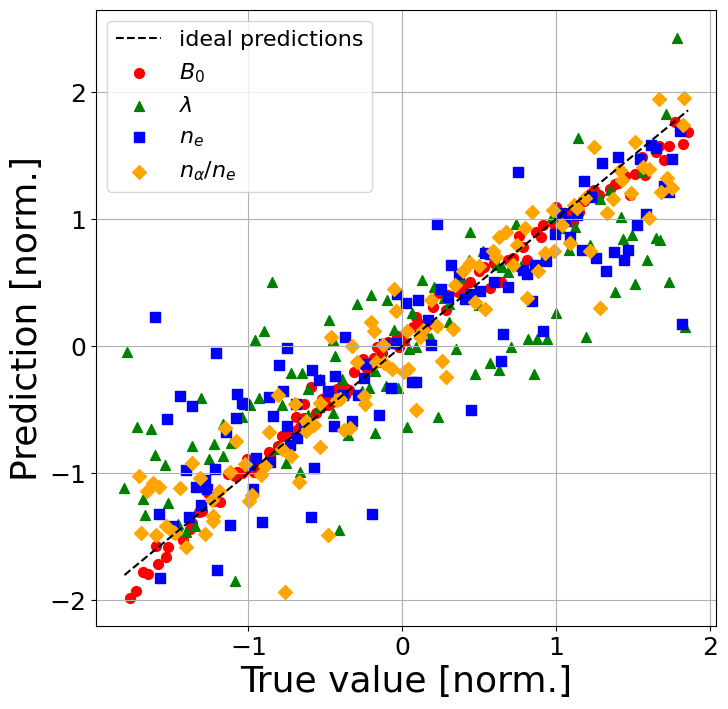}
    \caption{Hydra-generated normalised predictions of each parameter. The datapoints are the single test points left out in each fold of a leave-one-out cross-validation.}
\label{fig:individualPredictions}
\end{figure}

Our pipeline was run on 13 extrinsic regression algorithms available in Aeon: Catch22~\cite{lubba2019catch22}, Hydra~\cite{dempster2022hydracompetingconvolutionalkernels}, KNeighborsTimeSeries~\cite{JMLR:v25:23-1444}, MiniROCKET~\cite{dempster2021miniROCKET}, MultiROCKET~\cite{tan2021multiROCKET}, MultiROCKET-Hydra~\cite{dempster2022hydracompetingconvolutionalkernels}, QUANT~\cite{dempster2024quant}, RandomInterval~\cite{JMLR:v25:23-1444}, RandomIntervalSpectralEnsemble~\cite{lines2018time}, RDST~\cite{guillaume2022random}, TimeSeriesForest~\cite{deng2013time}, Summary~\cite{JMLR:v25:23-1444}, and the Dummy regressor, which predicts the mean of the data. They can handle multivariate data and tractably train models for the four fields with $110$ folds per algorithm required for a LOOCV. Hydra was the best performing algorithm by $R^2$ for two out of four simulation parameters and was within one standard error (by root mean squared error (RMSE)) of the top performance on all parameters. All predictions made by Hydra for the four simulation parameters are shown in Figure~\ref{fig:individualPredictions}. On a 13th generation 4.6GHz Intel i5-1335U, training of 110 folds of Hydra took approximately 2.5 CPU hours, while prediction of a single value took 100ms.

From experiments using the synthetic $\delta B_z$, $\delta E_x$ and $\delta E_y$ spectra across four propagation angles, we determine that Aeon's implementations of Hydra, MultiROCKET-Hydra (MRH; excepting pitch) and RDST (excepting alpha concentration) perform the best on prediction of all four simulation inputs to within statistical significance, as defined by one standard error from the mean prediction. RMSE has the desirable property of units equal to the original data, so prediction errors can be interpreted in terms of physical quantities. A summary of these is shown in Table~\ref{tab:denormed_tser_rmses}, where $B_0$, pitch, density and fast alpha concentration are predicted to within $0.6\sigma$ of the training data mean.
\begin{table}[h!]
    \begin{ruledtabular}
        \begin{tabular}{l|l|l|l}
            \textrm{Parameter}&
            \textrm{Best Model}&
            \textrm{RMSE}&
            \textrm{RMSE (denorm.)}\\
            \colrule
            $B_0$ & KNN & 0.025 & 0.028T\\
            $\lambda\,(v_\parallel/|v|)$ & Hydra & 0.589 & 0.160\\
            $n_e$ & MRH & 0.435 & $[2.20, 3.96]\times10^{19} m^{-3}$\\
            $n_\alpha/n_e$ & Hydra & 0.316 & $[6.17, 14.1]\times10^{-4}$\\
        \end{tabular}
    \end{ruledtabular}
    \caption{Best RMSE values and errors (ranges for log-sampled variables) achieved during extrinsic regression of power spectra against simulation input parameters, including denormalised RMSE values in original units. KNN is Aeon's implementation of a standard 1-nearest-neighbour algorithm. Note that for $B_0$, 11/13 algorithms produced statistically identical $R^2$ values of greater than 98\%.}
    \label{tab:denormed_tser_rmses}
\end{table}

\begin{figure}
    \centering
    \includegraphics[width=\linewidth]{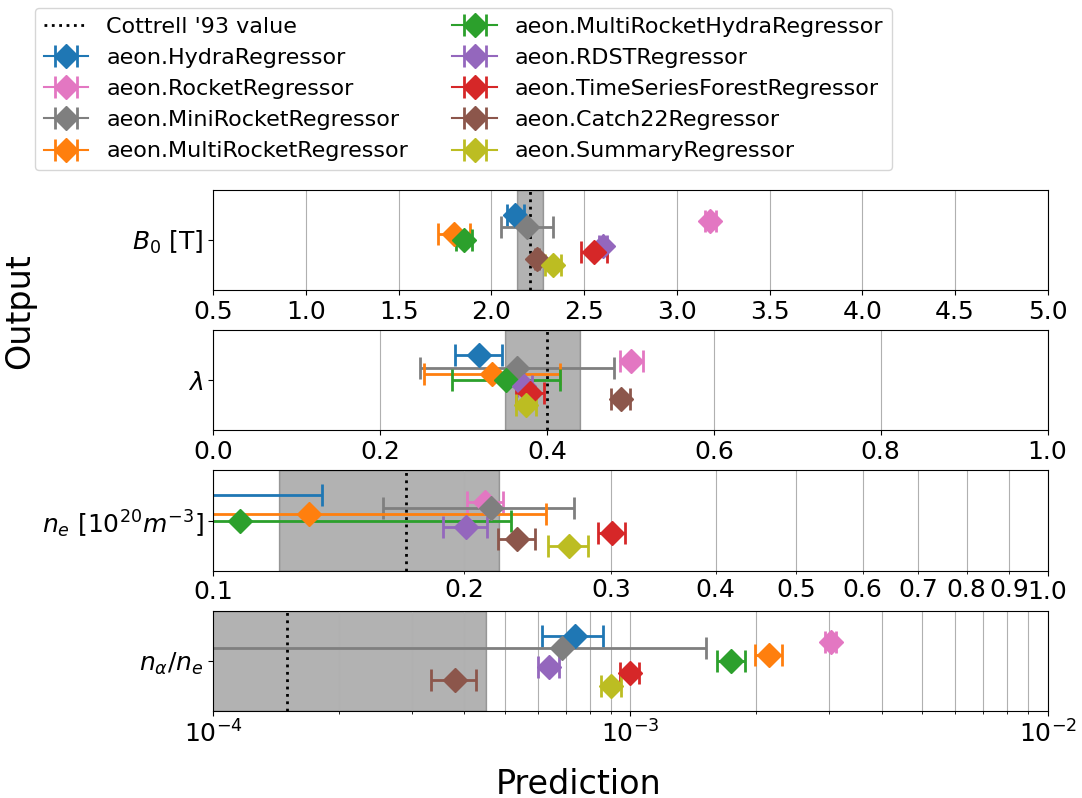}
    \caption{Predictions made by TSER models trained on $\delta B_z$ spectra from all simulations truncated to equal frequency bandwidths (${\sim}0-187$MHz) against JET pulse No. 26148 ICE spectrum data extracted from Ref.~\cite{cottrell1993ion}. Experimental values (excluding alpha-particle velocity pitch) are shown with dotted lines, with uncertainties denoted by shaded regions. The uncertainty in $B_0$ was estimated from measurements of the spectral peak spacing. The uncertainty in pitch was estimated from inspection of the shaded regions in Figures 15 and 18 of Ref.~\cite{cottrell1993ion}. The uncertainty in density was found from correlation with the range of possible major radius values in Figure 7~\cite{cottrell1993ion}. The uncertainty in alpha-particle concentration was based on rationalising the radial density profiles in Figure 7~\cite{cottrell1993ion} with estimates from TRANSP, and the large variation between core and edge values. The x-axis is scaled to include the entire training data simulation range.}
    \label{fig:cottrell}
\end{figure}

To provide a comparison to experiment, the ICE intensity spectrum from JET pulse No. 26148 was recovered from Figure 2 of Ref.~\cite{cottrell1993ion}. This JET shot used deuterium and 11\% tritium and was taken close to the time of peak neutron emission, supporting its use as an experimental comparison for ICE stimulated predominantly by fusion alphas in a bulk deuterium plasma. This spectrum resolved frequencies up to ${\sim}187$MHz and was obtained using a monopole antenna detecting the near-perpendicular component of $\vec{B}$. A comparable training set was built by truncating simulated $\delta B_z$ power spectra to this frequency range (resolving less than the first 5$\Omega_{c,\alpha}$ in the cases with smallest magnetic field). The unit of ICE intensity in the experimental data is dB without a stated physical baseline, so all simulated spectra were converted to a dB scale relative to their global power minimum ($2.23\times10^{-12}\mathrm{T}^2$). The ground truth value of $B = 2.21$T was evaluated based on the $17\pm0.5$MHz spacing of the spectral peaks. Alpha-particle velocity pitch of 0.4 was estimated from Figure 15 of Ref.~\cite{cottrell1993ion}. The density at the point of emission is inferred in Ref. ~\cite{cottrell1993ion} to be $1.7\times10^{19}\mathrm{m}^{-3}$. The alpha-particle fraction was evaluated by Cottrell et al. using TRANSP and stated in Figure 18 of Ref.~\cite{cottrell1993ion}. Training was run over all algorithms used during the previous LOOCV. Results from the nine best performing algorithms are shown in Figure~\ref{fig:cottrell}. The best results by mean absolute error per parameter are shown in Table~\ref{tab:cottrell_regression_results}.
\begin{table}[h!]
    \begin{ruledtabular}
        \begin{tabular}{l|r|r|r}
            \textrm{Parameter}&
            \textrm{True Value}&
            \textrm{Best Prediction}&
            \textrm{Best Model}\\
            \colrule
            $B_0$ [T] & $2.21\pm0.07$ & $2.19\pm0.14$ &MiniRocket\\
            $\lambda$ & $0.40\pm0.05$ & $0.38\pm0.02$ &TSF\\
            $n_e$ [$10^{19}m^{-3}$] & $1.7\pm0.5$ & $2.0\pm0.1$ &RDST\\
            $n_\alpha/n_e$ [$10^{-4}$] & $1.5\pm3.0$ & $3.8\pm0.5$ &Catch22\\
        \end{tabular}
    \end{ruledtabular}
    \caption{Values shown are the true experimental value, the best prediction by mean absolute error, and the model from which the best prediction was obtained. Uncertainty values for predictions are calculated as one standard deviation across 20 repeats. Uncertainty in the experimental data is transcribed or inferred from Ref.~\cite{cottrell1993ion}; see Figure~\ref{fig:cottrell}.}
    \label{tab:cottrell_regression_results}
\end{table}

A significant limitation when testing models against experimental data is the frequency bandwidth. The temporal length and output frequency of the simulations were set dynamically to ensure sufficient resolution up to 80$\Omega_{c,\alpha}$. In performing the LOOCV analysis, the TSER algorithms are passed spectra discretized onto an axis normalized to the per-simulation cyclotron frequency, and hence this normalization cannot provide discriminative frequency information. This information is added back by feeding in a fourth channel that maps cyclotron harmonics to Hz. When training the models for use with an experimental JET spectrum, however, the synthetic spectra are truncated and interpolated onto the frequency axis of the experimental observations. Hence higher frequency information is discarded; in the lowest-$B_0$ cases, the bandwidth is less than $5\Omega_{c,\alpha}$. 

Figure~\ref{fig:cottrell} shows an overlap between one standard deviation of MiniROCKET's prediction range and the experimental value's uncertainty for all parameters, with Hydra also predicting $B_0$ well, and MultiROCKET/MultiROCKET-Hydra working well for pitch and density to within uncertainty. Alpha concentration was generally predicted least successfully, though the experimental value is near the lower limit of the training data and most algorithms' predictions correctly return values lower than the training mean. Given the data range, collection and dB scale limitations making comparison to simulated spectra more difficult, this is a successful result. LOOCV performance indicates the capacity of a model to learn the training data, and the experimental comparison indicates the training data's proximity to a real spectrum and thus the predictive power of models trained only on synthetic spectra. 

The consistently highest performing models are from the convolution family of TSER algorithms. MultiROCKET-Hydra was one of the most accurate classifiers in a previous comparison in Ref.~\cite{middlehurst2024bake} of 40 time-series algorithms, and was almost 200 times faster to train than its comparably-accurate competition. This supports the hypothesis that training on power spectra from particle-in-cell codes is a suitable application for the time-series approach and bears similarity to standard time series classification benchmarks. Many of the regression algorithms in Aeon were adapted from classification variants~\cite{JMLR:v25:23-1444}. Hydra and the ROCKET-derivatives are recommended for generalisation.

Several caveats apply to these results. The periodic simulation domain implies a homogeneous, infinite 1D plasma without source or sink effects, complex geometry, spatially-varying background electromagnetic fields or mode conversion. 2D effects were incorporated by combining spectra from four angles near perpendicular, but given the qualitative differences observed even in these spectra separated by $2^\circ$, there is likely to be additional phenomena observed in spectra from diverse emission locations and subsequent propagation, attenuation and detection. Experimental spectra are likely to contain peaks affected by Doppler shifts~\cite{gorelenkov1995alfven}, doublet-splitting~\cite{cook2022doublet} and nonlinearities. These physical effects increase the complexity and fine structure of the spectra that arise from small differences in parameters that can be disambiguated by TSER algorithms. 

Regardless, Aeon's algorithms' success at predicting parameters based on an experimental spectrum, with all the real world physics it contains, having been trained only on simulated analogues is a credit to the robustness of the approach. A total data corpus of 110 simulations is small in machine learning terms for scanning four parameters simultaneously. The non-trivial inversion of this problem is made challenging by the covariance of ICE spectral characteristics with the various input parameters, which could be mitigated by increasing the dataset size and restricting the parameters to physically accessible regions of phase space. Improvements to a diagnostically-relevant ICE model could be achieved by: running fully kinetic 2D simulations in a restricted parameter space tailored to a particular device; increasing the training size as much as computational resources allow while scanning additional variables known to impact ICE such as temperature and $v_\perp/v_\mathrm{Alfv\acute{e}n}$~\cite{reman2019interpreting}; adding tritium to the background species; varying the background electron and fuel ion temperatures, potentially independently; and using a dynamic temporal simulation length to ensure the MCI fully saturates in all cases. It should also be noted that these particular plasma parameters considered herein were chosen based on both their operational usefulness and qualitative estimations of their likelihood to influence the MCI. In particular, if other information can disambiguate the up-down symmetry in a tokamak, $B_0$ and density information can spatially localise the point of emission. There may be many other quantities encoded in ICE spectra recoverable through this workflow. 

By combining machine learning and first-principles nonlinear kinetic plasma physics, this letter describes the creation of a control room-ready tool enabling the inference of bulk and fast ion plasma parameters from experimental ICE spectra for the first time. We ran 110 plasma parameter sets of nonlinear 1D3V fully-kinetic PIC simulations at four propagations angles near perpendicular to $B_0$ to build linear-in-2D composite spectra. These simulations comprise fusion alpha-particles in the drifting ``ring-beam'' class of distribution function with thermal fuel deuterons and electrons, a scenario known to produce spectra characteristic of ICE. This took only 1.5 million CPU hours, which enables most labs to generate comparable training sets and build similar inference models for their devices. TSER models were trained in a novel plasma physics context to predict $B_0$ strength (peak LOOCV $R^2=0.99$), electron density ($R^2=0.82$) and alpha-particle velocity pitch ($R^2=0.66$) and concentration ($R^2=0.90$) from synthetic $\delta B_z$, $\delta E_x$ and $\delta E_y$ frequency spectra derived from simulations. Comparison to a gold-standard JET ICE spectrum from Cottrell et al.~\cite{cottrell1993ion} required only one log-scale $\delta B_z$ spectrum truncated in frequency to $187$ MHz, well below today's diagnostic bandwidth capability. Despite these obstacles, the best performing algorithms were able to, based on the experimental spectrum only, predict the emitting plasma's $B_0$ to within 2\%, alpha velocity pitch to within 5\% and density to within 1\%. Without performance optimisation and with the largest training sets, the inference time for the best performing models (Hydra, MultiROCKET-Hydra) is around 100ms, making this approach suitable for inter-shot analysis of ICE spectra given a pre-trained prediction model. With conceivable performance optimisations, it could be extended to real-time control applications for the next generation of fusion power plants.

\section*{Acknowledgements}
This work was funded jointly by the UK Atomic Energy Authority and as part of the EPSRC Centre for Doctoral Training in Fusion Energy Science and Technology (Fusion Power CDT) with project reference EP/S022430/1. EPOCH was in part funded by the EPSRC grants EP/G054950/1, EP/G056803/1, EP/G055165/1, EP/ M022463/1 and EP/P02212X/1. This work used the ARCHER2 UK National Supercomputing Service (\url{https://www.archer2.ac.uk}) through EPSRC grant EP/X035336/1. The Viking cluster was also used during this project, which is a high performance compute facility provided by the University of York. We are grateful for computational support from the University of York, IT Services and the Research IT team.

\section*{Data Availability}
This work used EPOCH1D version 4.19.6 (commit \verb|f42ebadc|) available at \url{https://github.com/epochpic/epoch/tree/5.0-devel}. Relevant analysis code and data can be accessed at \url{https://doi.org/10.5281/zenodo.20122434}.

\bibliography{apssamp}

@inproceedings{todd2014diagnostic,
  title={Diagnostic systems in DEMO: engineering design issues},
  author={Todd, TN},
  booktitle={AIP Conference Proceedings},
  volume={1612},
  pages={9--16},
  year={2014},
  organization={American Institute of Physics}
}

@article{cottrell1988superthermal,
  title={Superthermal radiation from fusion products in JET},
  author={Cottrell, GA and Dendy, RO},
  journal={Physical review letters},
  volume={60},
  number={1},
  pages={33},
  year={1988},
  publisher={APS}
}

@article{chapman2020origin,
  title={Origin of ion cyclotron emission at the proton cyclotron frequency from the core of deuterium plasmas in the ASDEX-Upgrade tokamak},
  author={Chapman, B and Dendy, R O and Chapman, S C and McClements, K G and Ochoukov, R},
  journal={Plasma Physics and Controlled Fusion},
  volume={62},
  number={9},
  pages={095022},
  year={2020},
  publisher={IOP Publishing}
}

@article{reman2019interpreting,
  title={Interpreting observations of ion cyclotron emission from large helical device plasmas with beam-injected ion populations},
  author={Reman, Bernard Charles G and Dendy, R O and Akiyama, T and Chapman, Sandra C and Cook, J W S and Igami, H and Inagaki, S and Saito, K and Yun, GS},
  journal={Nuclear Fusion},
  volume={59},
  number={9},
  pages={096013},
  year={2019},
  publisher={IOP Publishing}
}

@article{carbajal2014linear,
  title={Linear and nonlinear physics of the magnetoacoustic cyclotron instability of fusion-born ions in relation to ion cyclotron emission},
  author={Carbajal, L and Dendy, R O and Chapman, Sandra C and Cook, James William S},
  journal={Physics of Plasmas},
  volume={21},
  number={1},
  year={2014},
  publisher={AIP Publishing}
}

@article{dendy1995ion,
  title={Ion cyclotron emission due to collective instability of fusion products and beam ions in TFTR and JET},
  author={Dendy, R O and McClements, Kenneth George and Lashmore-Davies, CN and Cottrell, GA and Majeski, R and Cauffman, S},
  journal={Nuclear fusion},
  volume={35},
  number={12},
  pages={1733},
  year={1995},
  publisher={IOP Publishing}
}

@article{cook2013particle,
  title={Particle-in-cell simulations of the magnetoacoustic cyclotron instability of fusion-born alpha-particles in tokamak plasmas},
  author={Cook, James William S and Dendy, R O and Chapman, Sandra C},
  journal={Plasma Physics and Controlled Fusion},
  volume={55},
  number={6},
  pages={065003},
  year={2013},
  publisher={IOP Publishing}
}

@article{tan2021time,
  title={Time series extrinsic regression: Predicting numeric values from time series data},
  author={Tan, Chang Wei and Bergmeir, Christoph and Petitjean, Fran{\c{c}}ois and Webb, Geoffrey I},
  journal={Data Mining and Knowledge Discovery},
  volume={35},
  number={3},
  pages={1032--1060},
  year={2021},
  publisher={Springer}
}

@article{Arber:2015hc,
  author = {Arber, T D and Bennett, K and Brady, C S and Lawrence-Douglas,
            A and Ramsay, M G and Sircombe, N J and Gillies, P and Evans,
            R G and Schmitz, H and Bell, A R and Ridgers, C P},
  title = {{Contemporary particle-in-cell approach to laser-plasma modelling}},
  journal = {Plasma Physics and Controlled Fusion},
  year = {2015},
  volume = {57},
  number = {11},
  pages = {1--26},
  month = nov
}

@article{middlehurst2024bake,
  title={Bake off redux: a review and experimental evaluation of recent time series classification algorithms},
  author={Middlehurst, Matthew and Sch{\"a}fer, Patrick and Bagnall, Anthony},
  journal={Data Mining and Knowledge Discovery},
  volume={38},
  number={4},
  pages={1958--2031},
  year={2024},
  publisher={Springer}
}

@article{wang2025time,
  title={Time series extrinsic regression for reconstructing missing electron temperature in tokamak},
  author={Wang, Minglong and Wan, Chenguang and Lu, Jingjing and Yu, Zhi and Xiao, Bingjia and Li, Yanlong and He, Xiaoxue and Luo, Zhengping and Yuan, Qiping and Hu, Yemin and others},
  journal={Nuclear Fusion},
  volume={65},
  number={7},
  pages={076008},
  year={2025},
  publisher={IOP Publishing}
}

@article{slade2024consequences,
  title={The consequences of tritium mix for simulated ion cyclotron emission spectra from deuterium-tritium plasmas},
  author={Slade-Harajda, Tobias Wieslaw and Chapman, Sandra C and Dendy, R O},
  journal={Nuclear Fusion},
  volume={64},
  number={12},
  pages={126051},
  year={2024},
  publisher={IOP Publishing}
}

@article{mcclements2018observations,
  title={Observations and modelling of ion cyclotron emission observed in JET plasmas using a sub-harmonic arc detection system during ion cyclotron resonance heating},
  author={McClements, Kenneth G and Brisset, Alexandra and Chapman, Benjamin and Chapman, Sandra C and Dendy, Richard O and Jacquet, Philippe and Kiptily, V G and Mantsinen, Mervi and Reman, Bernard C G and Contributors, JET},
  journal={Nuclear Fusion},
  volume={58},
  number={9},
  pages={096020},
  year={2018},
  publisher={IOP Publishing}
}

@article{mcclements1996interpretation,
  title={Interpretation of ion cyclotron emission from sub-Alfv{\'e}nic fusion products in the Tokamak Fusion Test Reactor},
  author={McClements, Kenneth George and Dendy, R O and Lashmore-Davies, C N and Cottrell, G A and Cauffman, S and Majeski, R},
  journal={Physics of Plasmas},
  volume={3},
  number={2},
  pages={543--553},
  year={1996},
  publisher={American Institute of Physics}
}

@article{cottrell1993ion,
  title={Ion cyclotron emission measurements during JET deuterium-tritium experiments},
  author={Cottrell, G A and Bhatnagar, V P and Da Costa, O and Dendy, R O and Jacquinot, J and McClements, K G and McCune, D C and Nave, M F F and Smeulders, P and Start, D F H},
  journal={Nuclear Fusion},
  volume={33},
  number={9},
  pages={1365},
  year={1993},
  publisher={IOP Publishing}
}

@book{birdsall1991plasma,
  title={Plasma physics via computer simulation},
  author={Birdsall, Charles K and Langdon, A Bruce},
  year={1991},
  publisher={CRC Press}
}

@book{hockney1988computer,
  title={Computer simulation using particles},
  author={Hockney, Roger W and Eastwood, James W},
  year={1988},
  publisher={CRC Press}
}

@article{gorelenkov2016ion,
  title={Ion cyclotron emission studies: retrospects and prospects},
  author={Gorelenkov, N N},
  journal={Plasma Physics Reports},
  volume={42},
  number={5},
  pages={430--439},
  year={2016},
  publisher={Springer}
}

@article{chapman2020comparing,
  title={Comparing theory and simulation of ion cyclotron emission from energetic ion populations with spherical shell and ring-beam distributions in velocity-space},
  author={Chapman, B and Dendy, R O and Chapman, S C and Holland, L A and Irvine, S W A and Reman, B C G},
  journal={Plasma Physics and Controlled Fusion},
  volume={62},
  number={5},
  pages={055003},
  year={2020},
  publisher={IOP Publishing}
}

@article{dendy2023mechanism,
  title={Mechanism for collective energy transfer from neutral beam-injected ions to fusion-born alpha particles on cyclotron timescales in a plasma},
  author={Dendy, R O and Chapman-Oplopoiou, B and Reman, B C G and Cook, J W S},
  journal={Physical Review Letters},
  volume={130},
  number={10},
  pages={105102},
  year={2023},
  publisher={APS}
}

@article{cook2022doublet,
  title={Doublet splitting of fusion alpha particle driven ion cyclotron emission},
  author={Cook, J W S},
  journal={Plasma Physics and Controlled Fusion},
  volume={64},
  number={11},
  pages={115002},
  year={2022},
  publisher={IOP Publishing}
}

@article{Sips_2005,
doi = {10.1088/0741-3335/47/5A/003},
url = {https://doi.org/10.1088/0741-3335/47/5A/003},
year = {2005},
month = {apr},
publisher = {},
volume = {47},
number = {5A},
pages = {A19},
author = {Sips, A C C and for the Steady State Operation and the Transport Physics topical groups of the International Tokamak Physics Activity},
title = {Advanced scenarios for ITER operation},
journal = {Plasma Physics and Controlled Fusion}
}

@article{Tholerus_2024,
doi = {10.1088/1741-4326/ad6ea2},
url = {https://doi.org/10.1088/1741-4326/ad6ea2},
year = {2024},
month = {aug},
publisher = {IOP Publishing},
volume = {64},
number = {10},
pages = {106030},
author = {Tholerus, E. and Casson, F.J. and Marsden, S.P. and Wilson, T. and Brunetti, D. and Fox, P. and Freethy, S.J. and Hender, T.C. and Henderson, S.S. and Hudoba, A. and Kirov, K.K. and Koechl, F. and Meyer, H. and Muldrew, S.I. and Olde, C. and Patel, B.S. and Roach, C.M. and Saarelma, S. and Xia, G. and the STEP team},
title = {Flat-top plasma operational space of the STEP power plant},
journal = {Nuclear Fusion}
}

@misc{beckett_2024_14507040,
  author       = {Beckett, George and
                  Beech-Brandt, Josephine and
                  Leach, Kieran and
                  Payne, Zöe and
                  Simpson, Alan and
                  Smith, Lorna and
                  Turner, Andy and
                  Whiting, Anne},
  title        = {ARCHER2 Service Description},
  month        = dec,
  year         = 2024,
  publisher    = {Zenodo},
  doi          = {10.5281/zenodo.14507040},
  url          = {https://doi.org/10.5281/zenodo.14507040},
}

@article{JMLR:v25:23-1444,
  author  = {Matthew Middlehurst and Ali Ismail-Fawaz and Antoine Guillaume and Christopher Holder and David Guijo-Rubio and Guzal Bulatova and Leonidas Tsaprounis and Lukasz Mentel and Martin Walter and Patrick Sch{{\"a}}fer and Anthony Bagnall},
  title   = {Aeon: a Python Toolkit for Learning from Time Series},
  journal = {Journal of Machine Learning Research},
  year    = {2024},
  volume  = {25},
  number  = {289},
  pages   = {1--10},
  url     = {http://jmlr.org/papers/v25/23-1444.html}
}

@inproceedings{bagnall2024hands,
  title={A hands-on introduction to time series classification and regression},
  author={Bagnall, Anthony and Middlehurst, Matthew and Forestier, Germain and Ismail-Fawaz, Ali and Guillaume, Antoine and Guijo-Rubio, David and Tan, Chang Wei and Dempster, Angus and Webb, Geoffrey I},
  booktitle={Proceedings of the 30th ACM SIGKDD Conference on Knowledge Discovery and Data Mining},
  pages={6410--6411},
  year={2024}
}

@article{guijo2024unsupervised,
  title={Unsupervised feature based algorithms for time series extrinsic regression},
  author={Guijo-Rubio, David and Middlehurst, Matthew and Arcencio, Guilherme and Silva, Diego Furtado and Bagnall, Anthony},
  journal={Data Mining and Knowledge Discovery},
  volume={38},
  number={4},
  pages={2141--2185},
  year={2024},
  publisher={Springer}
}

@inproceedings{middlehurst2023extracting,
  title={Extracting features from random subseries: A hybrid pipeline for time series classification and extrinsic regression},
  author={Middlehurst, Matthew and Bagnall, Anthony},
  booktitle={International Workshop on Advanced Analytics and Learning on Temporal Data},
  pages={113--126},
  year={2023},
  organization={Springer}
}

@misc{dempster2022hydracompetingconvolutionalkernels,
      title={HYDRA: Competing convolutional kernels for fast and accurate time series classification}, 
      author={Angus Dempster and Daniel F. Schmidt and Geoffrey I. Webb},
      year={2022},
      eprint={2203.13652},
      archivePrefix={arXiv},
      primaryClass={cs.LG},
      url={https://arxiv.org/abs/2203.13652}, 
}

@article{lubba2019catch22,
  title={Catch22: CAnonical Time-series CHaracteristics: Selected through highly comparative time-series analysis},
  author={Lubba, Carl H and Sethi, Sarab S and Knaute, Philip and Schultz, Simon R and Fulcher, Ben D and Jones, Nick S},
  journal={Data mining and knowledge discovery},
  volume={33},
  number={6},
  pages={1821--1852},
  year={2019},
  publisher={Springer}
}

@inproceedings{dempster2021minirocket,
  title={Minirocket: A very fast (almost) deterministic transform for time series classification},
  author={Dempster, Angus and Schmidt, Daniel F and Webb, Geoffrey I},
  booktitle={Proceedings of the 27th ACM SIGKDD conference on knowledge discovery \& data mining},
  pages={248--257},
  year={2021}
}

@article{tan2021multirocket,
  title={MultiRocket: multiple pooling operators and transformations for fast and effective time series classification},
  author={Tan, Chang Wei and Dempster, Angus and Bergmeir, Christoph and Webb, Geoffrey I},
  journal={arXiv preprint arXiv:2102.00457},
  year={2021}
}

@inproceedings{guillaume2022random,
  title={Random dilated shapelet transform: A new approach for time series shapelets},
  author={Guillaume, Antoine and Vrain, Christel and Elloumi, Wael},
  booktitle={International Conference on Pattern Recognition and Artificial Intelligence},
  pages={653--664},
  year={2022},
  organization={Springer}
}

@article{deng2013time,
  title={A time series forest for classification and feature extraction},
  author={Deng, Houtao and Runger, George and Tuv, Eugene and Vladimir, Martyanov},
  journal={Information Sciences},
  volume={239},
  pages={142--153},
  year={2013},
  publisher={Elsevier}
}

@article{dempster2024quant,
  title={Quant: A minimalist interval method for time series classification},
  author={Dempster, Angus and Schmidt, Daniel F and Webb, Geoffrey I},
  journal={Data Mining and Knowledge Discovery},
  volume={38},
  number={4},
  pages={2377--2402},
  year={2024},
  publisher={Springer}
}

@article{gorelenkov1995alfven,
  title={Alfv{\'e}n cyclotron instability and ion cyclotron emission},
  author={Gorelenkov, NN and Cheng, CZ},
  journal={Nuclear fusion},
  volume={35},
  number={12},
  pages={1743--1752},
  year={1995}
}

@article{gormezano2008chapter,
  title={Chapter 4: Advanced tokamak scenario development at jet},
  author={Gormezano, C and Challis, CD and Joffrin, E and Litaudon, X and Sips, ACC},
  journal={Fusion science and technology},
  volume={53},
  number={4},
  pages={958--988},
  year={2008},
  publisher={Taylor \& Francis}
}

@article{lines2018time,
  title={Time series classification with HIVE-COTE: The hierarchical vote collective of transformation-based ensembles},
  author={Lines, Jason and Taylor, Sarah and Bagnall, Anthony},
  journal={ACM Transactions on Knowledge Discovery from Data (TKDD)},
  volume={12},
  number={5},
  pages={1--35},
  year={2018},
  publisher={ACM New York, NY, USA}
}

@article{chapman2018nonlinear,
  title={Nonlinear wave interactions generate high-harmonic cyclotron emission from fusion-born protons during a KSTAR ELM crash},
  author={Chapman, Benjamin and Dendy, Richard O and Chapman, Sandra C and McClements, Kenneth G and Yun, Gunsu S and Thatipamula, Shekar Goud and Kim, MH},
  journal={Nuclear Fusion},
  volume={58},
  number={9},
  pages={096027},
  year={2018},
  publisher={IOP Publishing}
}

@article{dendy2015ion,
  title={Ion cyclotron emission from fusion-born ions in large tokamak plasmas: a brief review from JET and TFTR to ITER},
  author={Dendy, RO and McClements, KG},
  journal={Plasma Physics and controlled fusion},
  volume={57},
  number={4},
  pages={044002},
  year={2015},
  publisher={IOP Publishing}
}

@article{ochoukov2018observations,
  title={Observations of core ion cyclotron emission on ASDEX Upgrade tokamak},
  author={Ochoukov, R and Bobkov, V and Chapman, Benjamin and Dendy, R and Dunne, M and Faugel, H and Garc{\'\i}a-Mu{\~n}oz, Manuel and Geiger, B and Hennequin, Pascale and McClements, KG and others},
  journal={Review of Scientific Instruments},
  volume={89},
  number={10},
  year={2018},
  publisher={AIP Publishing}
}

@article{sato2010observation,
  title={Observation of ion cyclotron emission owing to DD fusion product H ions in JT-60U},
  author={Sato, Shoichi and Ichimura, Makoto and Yamaguchi, Yuusuke and Katano, Makoto and Imai, Yasutaka and Murakami, Tatsuya and Miyake, Yuichiro and Yokoyama, Takuro and Moriyama, Shinichi and Kobayashi, Takayuki and others},
  journal={Plasma and Fusion Research},
  volume={5},
  pages={S2067--S2067},
  year={2010},
  publisher={The Japan Society of Plasma Science and Nuclear Fusion Research}
}

@article{fulop1998ion,
  title={Ion cyclotron emission from fusion products and beam ions in the tokamak fusion test reactor},
  author={F{\"u}l{\"o}p, T{\"u}nde and Lisak, Mietek},
  journal={Nuclear fusion},
  volume={38},
  number={5},
  pages={761--773},
  year={1998}
}

@article{liu2020ion,
  title={Ion cyclotron emission driven by deuterium neutral beam injection and core fusion reaction ions in EAST},
  author={Liu, Lunan and Zhang, Xinjun and Zhu, Yubao and Qin, Chengming and Zhao, Yanping and Yuan, Shuai and Mao, Yuzhou and Li, Miaohui and Zhang, Kai and Cheng, Jian and others},
  journal={Nuclear Fusion},
  volume={60},
  number={4},
  pages={044002},
  year={2020},
  publisher={IOP Publishing}
}

@article{thome2019central,
  title={Central ion cyclotron emission in the DIII-D tokamak},
  author={Thome, Kathreen E and Pace, David C and Pinsker, Robert I and Van Zeeland, Michael A and Heidbrink, William W and Austin, Max E},
  journal={Nuclear Fusion},
  volume={59},
  number={8},
  pages={086011},
  year={2019},
  publisher={IOP Publishing}
}

@article{askinazi2018ion,
  title={Ion cyclotron emission in NBI-heated plasmas in the TUMAN-3M tokamak},
  author={Askinazi, LG and Belokurov, AA and Gin, DB and Kornev, VA and Lebedev, SV and Shevelev, AE and Tukachinsky, AS and Zhubr, NA},
  journal={Nuclear Fusion},
  volume={58},
  number={8},
  pages={082003},
  year={2018},
  publisher={IOP Publishing}
}

@article{mcclements2015fast,
  title={Fast particle-driven ion cyclotron emission (ICE) in tokamak plasmas and the case for an ICE diagnostic in ITER},
  author={McClements, KG and D'Inca, R and Dendy, RO and Carbajal, L and Chapman, Sandra C and Cook, James William S and Harvey, RW and Heidbrink, WW and Pinches, Simon David},
  journal={Nuclear Fusion},
  volume={55},
  number={4},
  pages={043013},
  year={2015},
  publisher={IOP Publishing}
}

@article{ochoukov2019interpretation,
  title={Interpretation of core ion cyclotron emission driven by sub-Alfv{\'e}nic beam-injected ions via magnetoacoustic cyclotron instability},
  author={Ochoukov, R and McClements, KG and Bilato, R and Bobkov, V and Chapman, B and Chapman, SC and Dendy, RO and Dreval, M and Faugel, H and Noterdaeme, J-M and others},
  journal={Nuclear Fusion},
  volume={59},
  number={8},
  pages={086032},
  year={2019},
  publisher={IOP Publishing}
}

@article{dendy1992possible,
  title={A possible excitation mechanism for observed superthermal ion cyclotron emission from tokamak plasmas},
  author={Dendy, RO and Lashmore-Davies, Chris N and Kam, KF},
  journal={Physics of Fluids B: Plasma Physics},
  volume={4},
  number={12},
  pages={3996--4006},
  year={1992},
  publisher={American Institute of Physics}
}

@article{fulop1997origin,
  title={Origin of superthermal ion cyclotron emission in tokamaks},
  author={Fulop, T and Kolesnichenko, Ya I and Lisak, Mietek and Anderson, Dan},
  journal={Nuclear fusion},
  volume={37},
  number={9},
  pages={1281--1293},
  year={1997}
}

@article{slade2025effects,
  title={Effects of deuterium-tritium mix on linear growth rates of the magnetoacoustic cyclotron instability in fusion plasmas},
  author={Slade-Harajda, Tobias Wieslaw and Cook, J W S and Dendy, Richard O and Chapman, Sandra C},
  journal={Physics of Plasmas},
  volume={32},
  number={8},
  year={2025},
  publisher={AIP Publishing}
}

@article{lanti2020orb5,
  title={ORB5: a global electromagnetic gyrokinetic code using the PIC approach in toroidal geometry},
  author={Lanti, Emmanuel and Ohana, No{\'e} and Tronko, Natalia and Hayward-Schneider, Thomas and Bottino, Alberto and McMillan, Ben F and Mishchenko, Alexey and Scheinberg, Aaron and Biancalani, Alessandro and Angelino, P and others},
  journal={Computer Physics Communications},
  volume={251},
  pages={107072},
  year={2020},
  publisher={Elsevier}
}

\end{document}